\documentstyle[hip-artc]{article}  

\volnumber{5}  \edyear{1997}  \frompage{000} \topage{000}                
\recrevdate{20 May, 2000}                                               
\def\rightmark{Nonlinear quantum amplifiers} 
\def\leftmark{A.\ Cs\'ot\'o, et al.}
 
\title{At the edge of nuclear stability: nonlinear \\ quantum amplifiers} 
\authors{
{\twerm Attila Cs\'ot\'o$^1$, Heinz Oberhummer$^2$, and Helmut Schlattl$^{3}$ %
}\\[2.812mm]
{\normalsize
\hspace*{-8pt}$^1$ Department of Atomic Physics, E\"otv\"os University, \\ 
P\'azm\'any P\'eter s\'et\'any 1/A, H-1117 Budapest, Hungary\\[0.2ex] 
\hspace*{-8pt}$^2$ Institute of Nuclear Physics, Vienna University 
of Technology,  \\ 
Wiedner Hauptstrasse 8-10, A-1040 Vienna, Austria \\[0.2ex] 
\hspace*{-8pt}$^3$ Max-Planck-Institut f\"ur Astrophysik, \\ 
Karl-Schwarzschild-Strasse 1, D-85741 Gar\-ching, Germany
}}
 
\abstract{We show that nuclei lying at the edge of stability can behave 
as nonlinear quantum amplifiers. A tiny change in the nucleon-nucleon 
interaction can trigger a much bigger change in the binding energy of 
these systems, relative to the few-cluster breakup threshold.}
 
\begin{document}
 
\maketitle
 
\section{Introduction}
 
Recently, the structure and reactions of nuclei far from stability have
received a lot of attention. Many of these systems possess interesting
properties, such as, e.g., halos, skins, etc. Most of these unusual features
are direct consequences of the very small binding energies of these nuclei, 
relative to the breakup into 2-3 clusters. We
would like to show a hitherto unnoticed interesting property of these systems:
at the edge of stability they may behave as nonlinear quantum amplifiers. 

Lying very close to the breakup threshold, the constituent clusters of a 
nucleus far from stability are barely held together by the residual forces 
between them. As the
energy moves closer to the breakup threshold, the nucleus becomes larger. The
rate at which the nucleus moves toward breakup, as a function of its size, is
determined by the rate at which the (attractive) residual interaction
between the clusters vanishes, as a function of their distance. One may imagine
situations (not necessarily in nuclear physics) where the residual
interaction drops to zero roughly as fast as the few-cluster binding 
energy, while the size
is increased. If this happens, then a tiny change in the basic interaction can
cause only a similarly small response in the binding energy. If, however, the
residual interaction between the clusters goes to zero much more slowly than
the binding energy does, then tiny changes in the basic interactions
can get substantially amplified in the binding energy \cite{amp}. We 
demonstrate this effect through the example of the $0^+_2$ state of 
$^{12}$C.

\section{The \boldmath $0^{\small \textbf{+}}_2$ state of $\mathbf{^{12}}$C as 
a quantum amplifier}

The $0^+_2$ state of $^{12}$C is famous for its role played in stellar
nucleosynthesis. Lying just 380 keV above the $3\alpha$ threshold,
this resonance is responsible for the synthesis of virtually all the carbon in
the Universe through a two-step capture of three alpha particles, the 
so-called triple-alpha process \cite{Rolfs}. Once $^{12}$C nuclei are produced, 
some of them are burned further in the 
$^{12}{\rm C}(\alpha,\gamma){^{16}}{\rm O}$ reaction to form $^{16}$O. 

We would like to see how sensitive the energy of the $0^+_2$ state is to the
fine details of the N-N interaction. For this purpose, we use a 12-body,
$3\alpha$-cluster model of $^{12}$C. The wave function of our model for 
$^{12}$C looks like
\begin{equation}
\label{wfn}
\Psi^{^{12}{\rm C}}=\sum_{l_1,l_2} {\cal A} \Bigl
\{ \Phi^\alpha \Phi^\alpha \Phi^\alpha\chi^{\alpha(
\alpha\alpha)}_{[l_1l_2]L} (\mbox{\boldmath$\rho$}_1,
\mbox{\boldmath$\rho$}_2) \Bigl\}.
\end{equation}
Here ${\cal A}$ is the intercluster antisymmetrizer, the $\Phi^\alpha$
cluster internal states are translationally invariant $0s$
harmonic-oscillator shell-model states, the
$\mbox{\boldmath$\rho$}$ vectors are the intercluster Jacobi
coordinates, $l_1$ and $l_2$ are the angular momenta of the two
relative motions, $L$ is the total orbital angular momentum, and
$[\ldots]$ denotes angular momentum coupling. 
Such a model was shown to give a good overall description of the low-lying
$^{12}$C states \cite{Pichler}. Here we follow the same method as in Ref.\
\cite{Pichler} to precisely determine the resonance energy of the $0^+_2$
state. 

In order to see the dependence of the results on the chosen effective
N-N interaction, we performed the calculations using four different
forces. The Minnesota (MN), Volkov 1 and 2 (V1, V2), 
and modified Hasegawa-Nagata (MHN) forces achieve similar quality in
describing light nuclear systems, including $^{12}$C \cite{Pichler}. We 
slightly adjusted a parameter (the
exchange mixture) of each forces in order to get the $0^+_2$ resonance
energy right at the experimental value. The results coming from these forces 
are our baseline predictions. Then, we multiplied the strengths of each forces  
by a factor $p$, which varied from $p=0.996$ to 1.004, and calculated the 
resonance energies again. This way we can monitor the response of the $0^+_2$ 
state to slight perturbations in the N-N strengths. 

The results for the resonance energies are 
shown in Table \ref{tab1}. 
\begin{table}[!t]
\vspace*{-12pt}
\caption[]{The energy $\varepsilon$ (in keV) of the $0^+_2$ resonance, relative
to the $3\alpha$ threshold, as a function of the strength factor $p$ of the 
various N-N interactions}
\begin{center}
\begin{minipage}{.8\textwidth}
\renewcommand{\footnoterule}{\kern -3pt} %
\begin{tabular}{ccccc}
\hline\\[-10pt]
 & \multicolumn{4}{c}{N-N interaction} \\
$p$ & MHN & MN & V1 & V2  \\
\hline\\[-10pt]
1.004 & 235.6 & 273.3 & 294.4 & 306.0 \\
1.002 & 308.1 & 327.5 & 337.5 & 343.7 \\
1.001 & 344.4 & 353.7 & 358.7 & 361.7 \\
1.000 & 379.6 & 379.6 & 379.6 & 379.6 \\
0.999 & 414.3 & 405.2 & 400.3 & 397.2 \\
0.998 & 448.8 & 430.5 & 420.8 & 414.6 \\
0.996 & 517.0 & 481.4 & 460.7 & 450.0 \\
\hline
\end{tabular}
\end{minipage}
\renewcommand{\footnoterule}{\kern-3pt \hrule width .4\columnwidth \kern 2.6pt}
\end{center}
\label{tab1}
\vspace*{-0.5cm}
\end{table}
As one can see, very small changes in the N-N interaction
can cause almost two orders of magnitude bigger changes in the resonance
energy. We believe that this is the consequence of the nonlinear amplification
phenomenon discussed above. Table \ref{tab1} shows that the different N-N 
interactions give rather different results for the dependence of the resonance 
energy $\varepsilon$ on $p$. Simultaneous studies of the $^8$Be ground state 
and the $0^+_2$ state of $^{12}$C indicate that the $\alpha-\alpha$ residual 
interaction is somewhat too strong in the case of the MHN interaction, while 
it is somewhat too weak in the case of the MN, V1, and V2 forces (increasingly 
weaker from MN to V2). Thus, we expect that the true behavior of $\varepsilon$ 
with respect to small perturbations in the N-N force (as shown in Table 
\ref{tab1}) is somewhere between the MHN and MN results. 

\section{Conclusions}

\begin{figure}[!t]
\vspace*{-0.1cm}
\begin{center}
\leavevmode
\hbox to\hsize{\hss\epsfxsize=330bp\epsfbox{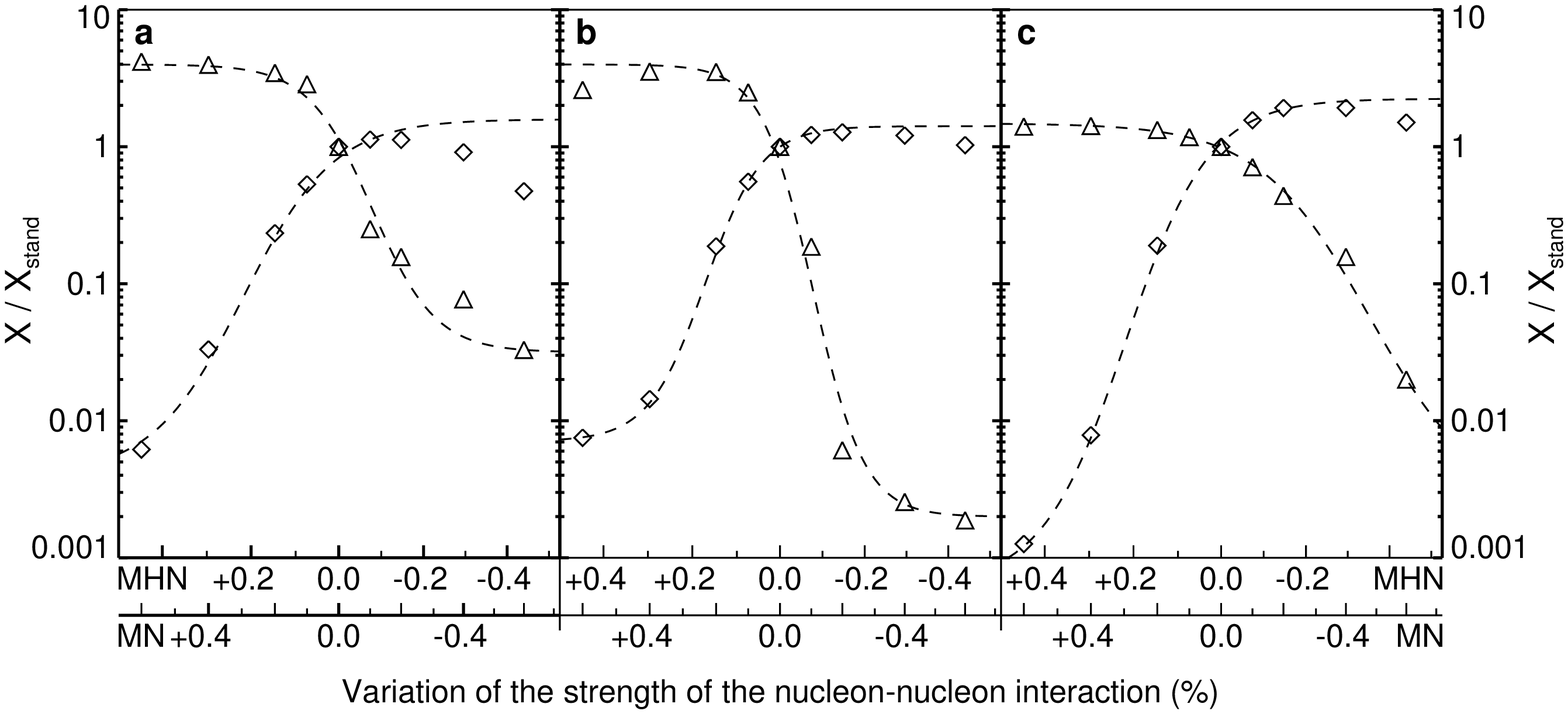}\hss} 
\end{center}
\vspace*{-0.7cm}
\caption[]{The change of the carbon ($\triangle$) and oxygen
($\Diamond$) mass abundances ($X$) through variations of the strength 
of the strong interaction. They are shown in panels a, b, and c for 
stars with masses of 20, 5, and $1.3M_\odot$, respectively, in units 
of the standard values $X_{\rm stand}$. The variations of the strength 
of the strong interaction are given for the two effective N-N forces MHN 
and MN. The dashed curves are drawn to guide the eye.
}
\label{fig1}
\vspace*{-0.5cm}
\end{figure}

The strong sensitivity of the $0^+_2$ resonance energy to small changes in the
N-N interaction has a spectacular consequence on the stellar production of 
carbon and oxygen. Since both steps of the triple-alpha reaction are governed 
by narrow resonances, the triple-alpha rate can be given \cite{Rolfs}, in a 
very good approximation, as 
\begin{equation}
\label{3alpha}
r_{3\alpha} = 3^{\frac{3}{2}} N_{\alpha}^3
\left(\frac{2 \pi \hbar^2}{M_{\alpha} k_{\rm B} T}\right)^3
\frac{\omega \gamma}{\hbar} \exp \left(-\frac{\varepsilon}{k_{\rm B} T}\right),
\end{equation}
where $M_{\alpha}$ and $N_{\alpha}$ is the mass and the number density of the
$\alpha$ particle, respectively, $T$ is the temperature of the stellar plasma, 
$\varepsilon$ is the resonance energy of the $0^+_2$ state, relative to the 
$3\alpha$ threshold, and $\omega \gamma$ is the resonance
strength. Since $r_{3\alpha}$ depends exponentially on $\varepsilon$, even
small variations in $\varepsilon$ can cause large changes in $r_{3\alpha}$. We
calculated the $r_{3\alpha}$ rates for the resonance energies shown in Table
\ref{tab1}, and used these rates in a contemporary stellar model code
\cite{Helmut} to see how much the tiny changes in the N-N force can influence
the synthesis of carbon and oxygen \cite{prev}. We performed calculations for 
low-mass, medium-mass, and massive stars. The resulting carbon and oxygen
abundances, relative to the standard abundances, are shown in Fig.\ \ref{fig1}
as a function of the change in the strength of the MHN and MN forces. 
One can easily understand the qualitative behavior of the results by noticing
that a stronger force leads to a smaller $\varepsilon$, thus to a larger
$r_{3\alpha}$, which results in a more effective triple-alpha process, relative
to the $^{12}{\rm C}(\alpha,\gamma){^{16}}{\rm O}$ burning. A similar but
reversed reasoning holds for a weaker force.

The really spectacular feature of Fig.\ \ref{fig1} is that the tiny changes in
the N-N force can cause enormous changes in the carbon and oxygen abundances.
At the very root of this behavior lies the nonlinear amplification 
phenomenon discussed above. We can say that a 0.5\% change in the N-N force
would lead to a situation where there is virtually no carbon or oxygen present,
which would make carbon-based life impossible. As the strength of the N-N force
is connected, through the pion mass, to the quark masses and ultimately to 
the vacuum expectation value of the Higgs field, our results can in principle
be used to give constraints on some fundamental parameters of the Standard
Model \cite{Jeltema}.

\section*{Acknowledgement}
This work was partly supported by D32513/FKFP-0242-2000/BO-00520-98 
(Hungary) and by the John Templeton Foundation 
(938-COS153).

\vfill\eject
\end{document}